\pdfoutput=1

\documentclass[twocolumn,prl,showpacs,amsmath,amssymb,superscriptaddress]{revtex4}

\usepackage{bm}
\usepackage{graphicx}
\usepackage{amssymb}
\usepackage{xcolor}

\begin{document}

\title{Multiband superconductivity in the correlated electron filled skutterudite system Pr$_{1-x}$Ce$_x$Pt$_4$Ge$_{12}$  }

\author{Y. P. Singh}  
\altaffiliation[Present Address: ]{Department of Mechanical Engineering, The University of Akron, Akron, Ohio, 44325, USA}
\affiliation{Department of Physics, Kent State University, Kent, Ohio, 44242, USA}

\author{R. B. Adhikari}  
\affiliation{Department of Physics, Kent State University, Kent, Ohio, 44242, USA}

\author{S. Zhang}  
\affiliation{Department of Physics, Kent State University, Kent, Ohio, 44242, USA}

\author{K. Huang}
\altaffiliation[Present Address: ]{Department of Physics, Fudan University, Shanghai, China, 200433}
\affiliation{Center for Advanced Nanoscience, University of California, San Diego, La Jolla, California 92093, USA}
\affiliation{Materials Science and Engineering Program, University of California, San Diego, La Jolla, California 92093, USA}

\author{D. Yazici}
\altaffiliation[Present Address: ]{Faculty of Health Sciences, Artvin Coruh University, Artvin 08100, Turkey}
\affiliation{Center for Advanced Nanoscience, University of California, San Diego, La Jolla, California 92093, USA}
\affiliation{Department of Physics, University of California at San Diego, La Jolla, CA 92903, USA}

\author{I. Jeon}
\affiliation{Center for Advanced Nanoscience, University of California, San Diego, La Jolla, California 92093, USA}
\affiliation{Materials Science and Engineering Program, University of California, San Diego, La Jolla, California 92093, USA}

\author{M. B. Maple}
\affiliation{Center for Advanced Nanoscience, University of California, San Diego, La Jolla, California 92093, USA}
\affiliation{Materials Science and Engineering Program, University of California, San Diego, La Jolla, California 92093, USA}
\affiliation{Department of Physics, University of California at San Diego, La Jolla, CA 92903, USA}

\author{M. Dzero}
\affiliation{Department of Physics, Kent State University, Kent, Ohio, 44242, USA}

\author{Carmen C. Almasan}
\affiliation{Department of Physics, Kent State University, Kent, Ohio, 44242, USA}

\date{\today}
\pacs{71.10.Hf, 71.27.+a, 74.70.Tx}

\begin{abstract}
Studies of superconductivity in multiband correlated electronic systems has become one of the central topics in condensed matter/materials physics.  In this paper, we present the results of thermodynamic measurements on the superconducting filled skutterudite system Pr$_{1-x}$Ce$_x$Pt$_4$Ge$_{12}$ ($ 0 \leq x \leq 0.2$) to investigate how substitution of Ce at Pr sites affects superconductivity. We find that an increase in Ce concentration leads to a suppression of the superconducting transition temperature from $T_{c}\sim 7.9$ K for $x=0$ to $T_c\sim 0.6$ K for $x=0.14$. Our analysis of the specific heat data for $x\leq 0.07$ reveals that superconductivity must develop in at least two bands: the superconducting order parameter has nodes on one Fermi pocket and remains fully gapped on the other. Both the nodal and nodeless gap values decrease, with the nodal gap being suppressed more strongly, with Ce substitution. Ultimately, the higher Ce concentration samples ($x>0.07$) display a nodeless gap only. 
\end{abstract}

\pacs{71.10.Ay, 74.25.F-, 74.62.Bf, 75.20.Hr}

\maketitle
\section{Introduction}
Filled skutterudite compounds with the chemical formula $M$Pt$_4$Ge$_{12}$ ($M =$ alkaline earth, lanthanide, or actinide) are a relatively new entry into the family of heavy-fermion superconductors. The first Pr-based heavy-fermion superconductor PrOs$_4$Sb$_{12}$ has a superconducting critical temperature $T_c\simeq 1.85$ K and a Sommerfeld coefficient $\gamma\sim 500$ mJ/(mol$\cdot$K$^2$), revealing a rather significant enhancement of the effective mass of the conduction electrons~\cite{Bauer2002,Maple2006}.  Interestingly, the related compound PrPt$_4$Ge$_{12}$ has a much higher $T_c\simeq{7.9}$ K and smaller $\gamma\sim 60$ mJ/(mol$\cdot$K$^2$), corresponding to a moderate enhancement of the conduction electron effective mass~\cite{Maisuradze2009}.   
It is noteworthy that these materials, among many others in the family of filled skutterudites, have been widely studied recently, mainly due to their potential for thermoelectric applications \cite{Sales1996, Keppens1998} as well as a variety of low temperature phenomena such as magnetic and quadrupolar order and metal-insulator transitions \cite{Aoki2006,Cichorek2005,Bauer2002,Maple2003,Maple2009,Uher2000}. 

The microscopic nature of superconductivity in both PrPt$_4$Ge$_{12}$ and PrOs$_4$Sb$_{12}$ has been the focus of intense experimental effort in recent years. For example, $\mu$SR experiments on PrPt$_4$Ge$_{12}$ reveal time reversal symmetry breaking (TRSB) in the superconducting state, indicating an unconventional symmetry of the pairing amplitude \cite{Zhang2015}. 
The substitution of Ce for Pr results in the suppression of the TRSB. In addition, the  specific heat of the systems Pr$_{1-x}$Ce$_x$Pt$_4$Ge$_{12}$ and PrPt$_4$Ge$_{12-x}$Sb$_x$ shows a crossover from a power law to an exponential temperature dependence upon increase of the Ce and Sb substituent concentration, which unequivocally suggests that the superconducting gap becomes fully isotropic at sufficiently high Ce or Sb concentrations \cite{Huang2014, Jeon2016}. 

Naively, one would not regard these observations as surprising, since several experimental probes have confirmed that
the parent compound PrPt$_4$Ge$_{12}$ is a multiband superconductor \cite{Maisuradze2009,Zhang2013}.
Given that cerium ions have a valence of 3+  ($4f^1$ configuration) corresponding to an odd number of electrons in the 4$f$-shell, one would generally expect a rapid  suppression of unconventional superconductivity in response to magnetic scattering~\cite{Huang2016}.
If the two Fermi pockets are not related by crystal symmetry, then the unconventional pairing on one of the Fermi pockets will be more susceptible to magnetic scattering than its conventional (or symmetric) counterpart. 
Although the viability of this interpretation still  
remains an open issue, the scenario of co-existence between nodal and nodeless pairing gaps finds support in different experimental findings \cite{Maisuradze2009,Zhang2013,Huang2014}. In that regard, it is worth mentioning that the same issues persist for PrOs$_4$Sb$_{12}$, for which several experimental groups have provided evidence for nodal as well as nodeless superconductivity \cite{Maple2006,Maple2009,Sato2010}.   
 
Motivated by these questions, we performed low-temperature specific heat  measurements on superconducting samples of Pr$_{1-x}$Ce$_x$Pt$_4$Ge$_{12}$. Our detailed and systematic analysis indicates the presence of multiband superconductivity up to the Ce concentration $x = 0.07$ and a single nodeless gap for $x > 0.07$. Furthermore, we argue that at least one of the superconducting order parameters for the parent compound and the lightly doped compound must be nodal. Thus, our findings imply that the nodal gap disappears at high Ce concentrations. Our findings are consistent with the suppression of the TRSB with increasing Ce concentration in these materials, as observed in recent $\mu$SR experiments \cite{Zhang2015}. 

\section{Experimental Details}
Polycrystalline  samples  of  Pr$_{1-x}$Ce$_x$Pt$_4$Ge$_{12}$ were synthesized by arc-melting and annealing, according to a procedure described in detail elsewhere \citep{Huang2014}. The surfaces of the samples were polished with sand paper to get better contact between the sample and the specific heat platform.  
We performed a series of specific heat measurements on the polycrystalline samples of Pr$_{1-x}$Ce$_x$Pt$_4$Ge$_{12}$ ($x =$ 0, 0.01, 0.03, 0.04, 0.05,  0.06,  0.07,  0.085, 0.1, and 0.14) in zero magnetic field and over the temperature $T$ range $0.50$ K $\leq T \leq 10$ K. The specific  heat  measurements  were  performed  via a standard thermal relaxation technique using the He-3 option of the Quantum Design's Physical Property Measurement System (PPMS). 

\section{Results and Discussion}
\begin{figure}
\centering
\includegraphics[width=1.0\linewidth]{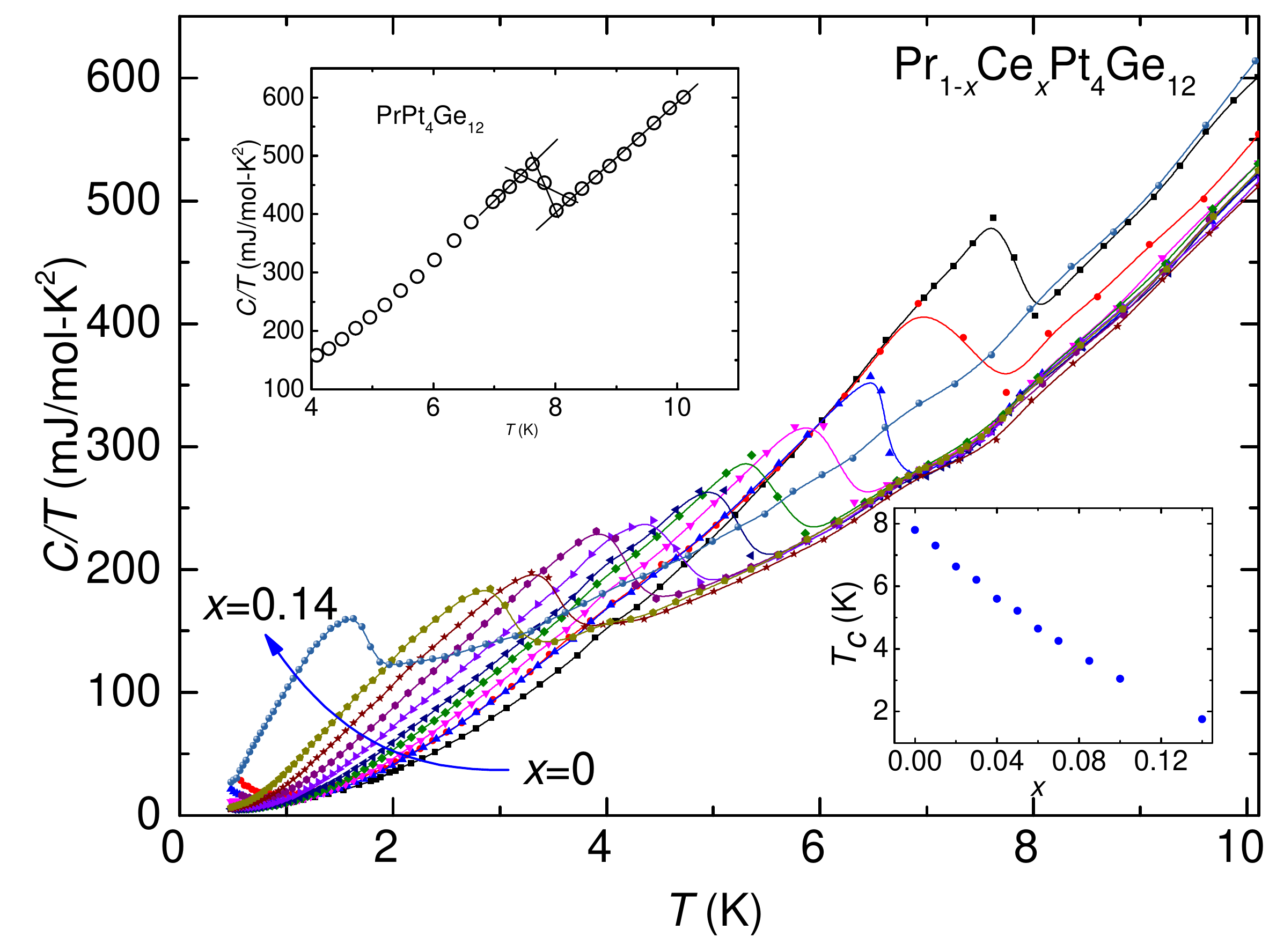}
\caption{(Color online) Measured specific heat $C$ divided by temperature $T$ plotted vs. $T$ for different Ce substitutent concentrations in Pr$_{1-x}$Ce$_x$Pt$_4$Ge$_{12}$ ($0 \leq x \leq 0.14$). The measurements were performed over the temperature range $0.50 \leq T \leq 10$ K. Top left inset: Isoentropic construction to obtain the value of superconducting transition temperature $T_c$.  Bottom right inset: Superconducting transition temperature $T_c$, obtained from the data in the main panel of the figure using the isoentropic construction, vs. $x$.}
\label{Fig1}
\end{figure}

As shown in Fig.~\ref{Fig1}, a clear superconducting transition is observed for Pr$_{1-x}$Ce$_x$Pt$_4$Ge$_{12}$ samples with $x \leq$ 0.14. Samples with $x >$ 0.14 do not display a complete superconducting transition for temperatures as low as 0.5 K. For the unambiguous determination of the thermodynamic superconducting transition temperature $T_c$ we used the method of \emph {isoentropic} construction; i.e., we chose $T_c$ such that the entropy around the transition is conserved (see top left inset to Fig.~\ref{Fig1}. The bottom right inset to Fig.~\ref{Fig1} shows that $T_c$ is suppressed monotonically with increasing Ce substitution $x$. 

\begin{figure}
\centering
\includegraphics[width=1\linewidth]{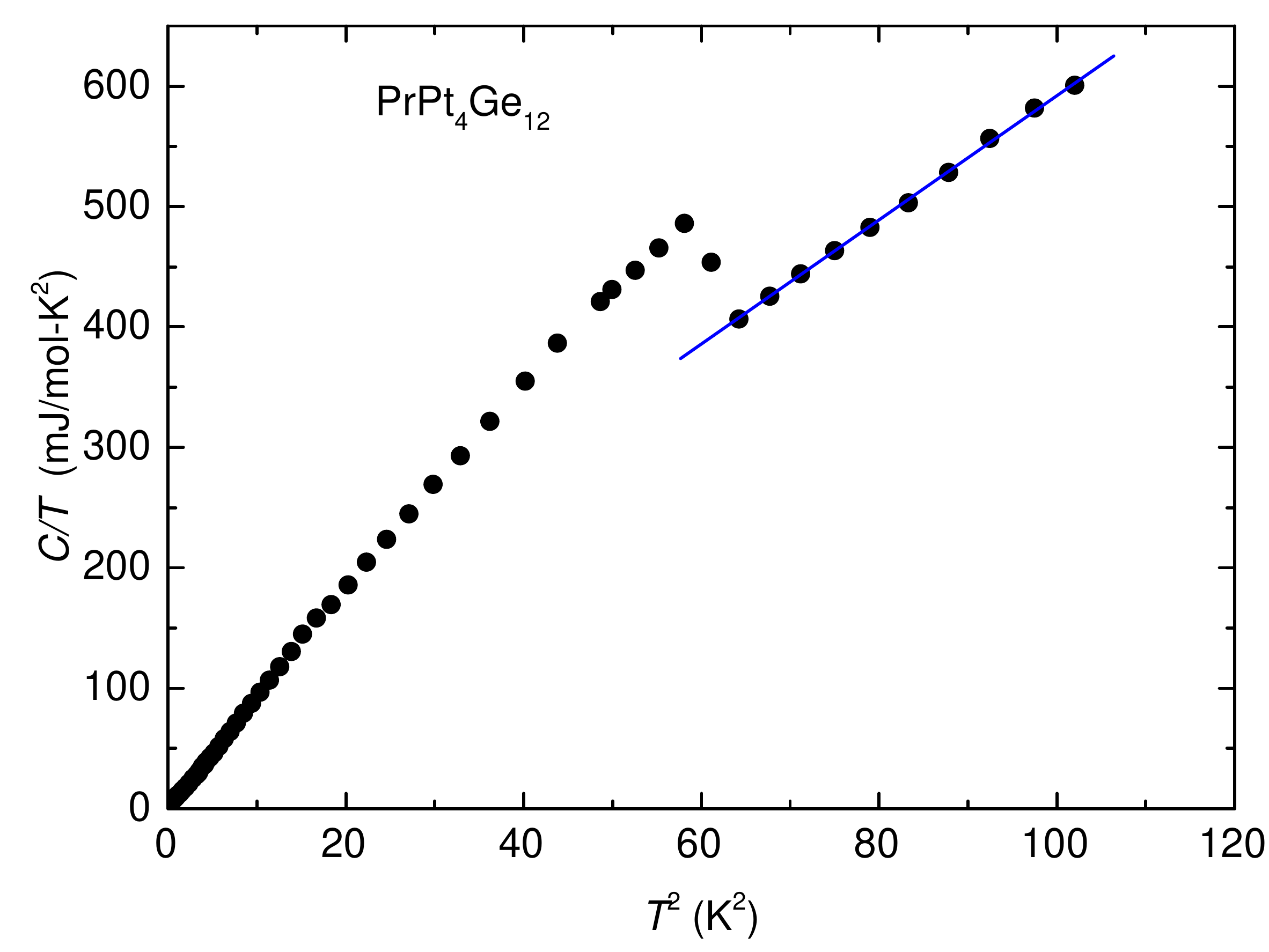}
\caption{ (Color online) (a) Fit (blue line) of the $C(T)/T$ data (solid filled circles) with $\gamma _n+ \beta T^2$ in the temperatures range just above $T_c$ to 10 K, performed in order to extract the lattice contribution to specific heat. The fit yields the parameters $\gamma_n = 73.7 $mJ/mol$\cdot$K$^2$ and $\beta = 5.2 $mJ/mol$\cdot$K$^4$.}
\label{Fig2}
\end{figure}

In the absence of any magnetic contribution, the measured specific heat in the normal state is the sum of electronic 
$C_{\rm e}=\gamma_n T$ and lattice $C_{\textrm{ph}} = \beta T^3$ contributions; hence, we fitted the measured specific heat in the normal state ($T_c < T \leq $ 10 K) for different Ce concentrations with $C(T)$ = $\gamma_n T$+$\beta T^3$. The result of such a fit for the  $x = 0$ sample is shown in Fig.~\ref{Fig2} and gives $\gamma_n = 73.7$ mJ/mol$\cdot$K$^2$ and $\beta = 5.2 $ mJ/mol$\cdot$K$^4$. To determine the electronic contribution to the specific heat, we subtracted the lattice contribution $C_{\textrm{ph}}$ from the measured specific heat.

\begin{figure}
\centering
\includegraphics[width=\linewidth]{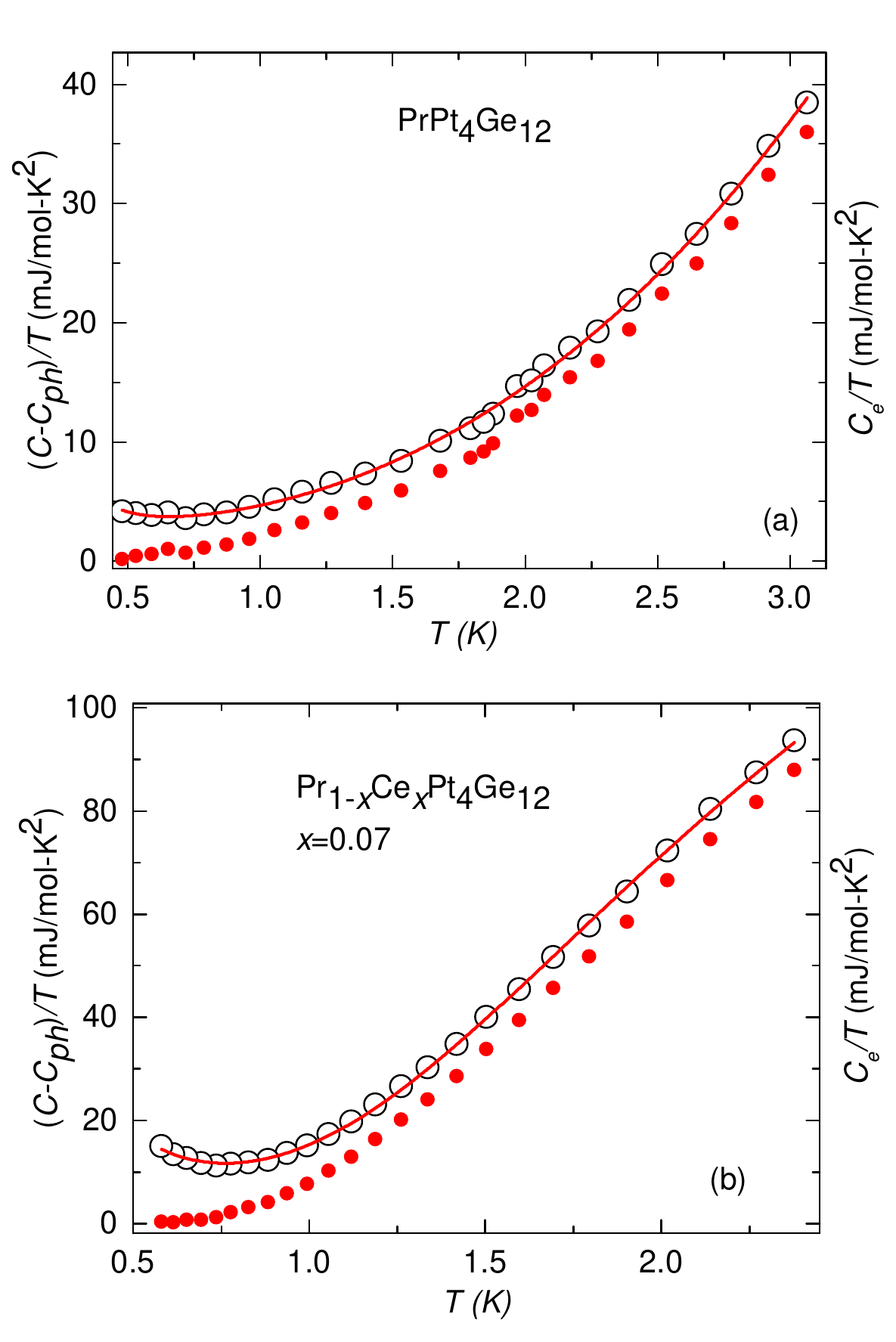}
\caption{ (Color online) Plot of the specific heat data $(C-C_{ph})/T$ vs temperature $T$ (black circles) for the (a) \textit{x} = 0 and (b) \textit{x} = 0.07 samples. The solid red lines are fits of the data as described in the text. The solid red circles represent the electronic contribution to specific heat (right vertical axis), obtained after subtraction of all the other contributions.}
\label{Fig3}
\end{figure}

After subtracting the lattice contribution, the specific heat shows an upturn in the low-temperature region (see Fig.~\ref{Fig3}), which we attribute to a nuclear Schottky contribution $C_{\textrm{Sch}} =  A_n/T^2$ due to impurity phases that appear during preparation of the samples.  In fact, such an upturn was also observed in previous studies \cite{Maisuradze2009,Zhang2013} and was suppressed in the case of single crystals by properly etching the surface of the crystals. Since the present study was performed on polycrystalline samples, we were unable to eliminate the nuclear Schottky contribution by just etching the surface of the samples. Nevertheless, the upturn in the present study appears randomly throughout the samples with different Ce concentration, leading us to conclude that it is, indeed, the result of impurity phases. Therefore, to determine the electronic contribution to specific heat, we also subtracted the nuclear Schottky contribution $C_{\textrm{Sch}}=A_n/T^2$ for temperatures $T < 2$ K where the effect of the upturn is evident. We obtained the best fit  of the data with the expression $(C-C_{ph})=\gamma_0T + BT^m + A_n/T^2$  for $ 0 \leq x \leq 0.04$ and with $(C-C_{ph})=\gamma_0T + Be^{-\Delta/T} + A_n/T^2$ for $x > 0.04$.  Here $\gamma_0$ and $A_n/T^2$ are the residual specific heat and the nuclear Schottky contribution, respectively. Figures~\ref{Fig3}(a) and \ref{Fig3}(b) show the results of the fitting (red curves) and the electronic contribution to the specific heat (red solid symbols) for the $x = 0$ and $x = 0.07$ sample, respectively, after subtracting the residual heat capacity and the nuclear Schottky contribution.

As we have already mentioned in the introduction, there are conflicting reports regarding the symmetry of the superconducting energy gap of PrPt$_4$Ge$_{12}$. For example, a specific heat study based on the BCS model points towards multiband superconductivity with isotropic order parameters \cite{Zhang2013}, while power-law behavior in the specific heat at low temperatures, together with measurement of the superfluid density through $\mu$-SR measurements, indicate a gap with point nodes \cite{Maisuradze2009}. This latter finding is also supported by the time-reversal symmetry breaking observed through another independent $\mu$-SR study, suggesting the superconductivity is of unconventional nature \cite{Zhang2015}.

\begin{figure}
\centering
\includegraphics[width=1\linewidth]{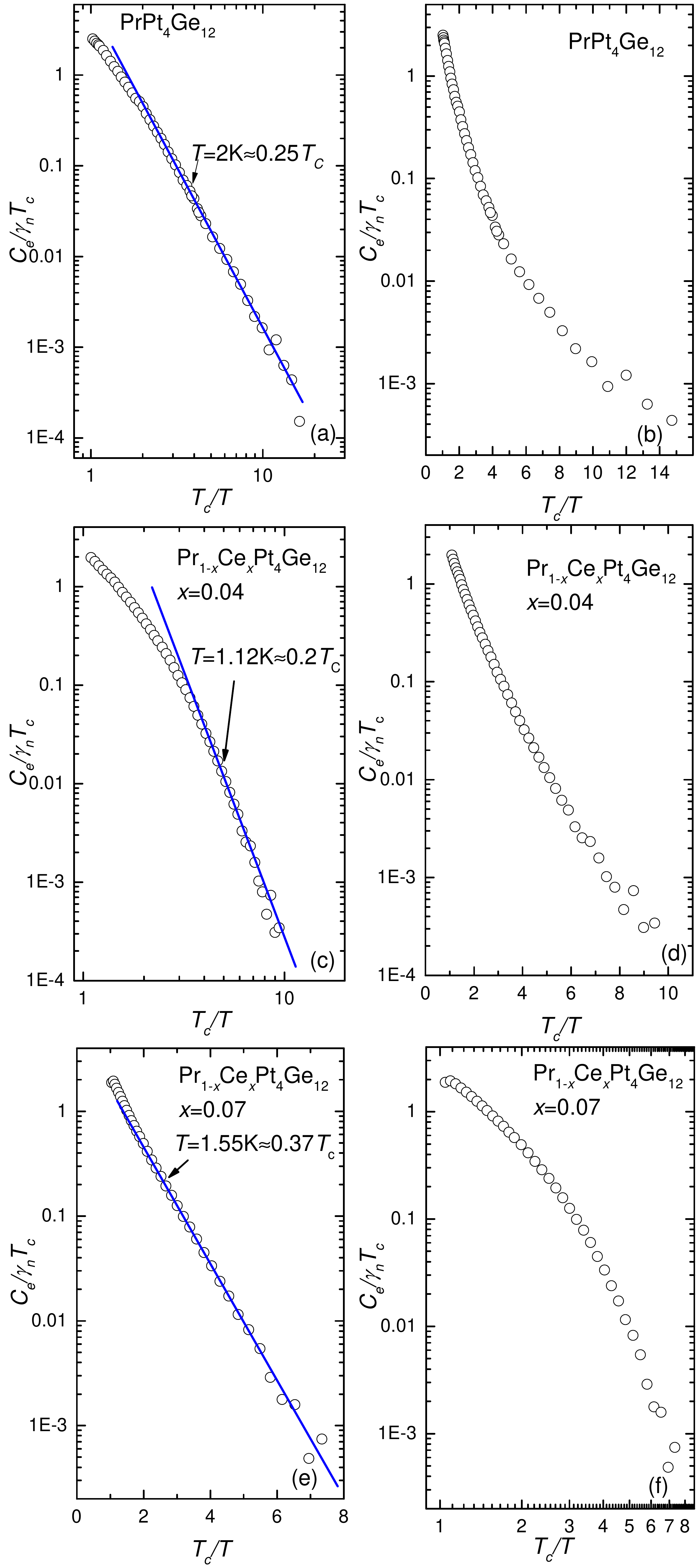}
\caption{(Color online) Log-log plots of $C_e /\gamma_nT_c$ vs. $T_c/T$ for (a) $x = 0$, (c) $x = 0.04$, and (f) $x = 0.07$ Ce concentrations. The blue lines show the temperature range over which the power law holds for $x = 0$ and $x = 0.04$ and the exponential works for $x = 0.07$.  Semi-log plots of $C_e /\gamma_nT_c$ vs. $T_c/T$ for (b) $x = 0$, (d) $x = 0.04$ and (e) $x = 0.07$ samples.}
\label{Fig4}
\end{figure}

To get better insight into the nature of the superconducting gap of this system and its evolution with Ce substitution, we first focus on the temperature dependence of the electronic specific heat at low $T$. As just mentioned, the power-law $T$ dependence of the electronic specific heat at low temperatures ($T\ll T_c$) is a signature of the nodal character of the superconducting order parameter \cite {Sigrist1991}, while within the analysis based on the weak-coupling BCS theory, its exponential temperature dependence characterizes an isotropic gap. Thus, in our analysis, we use log-log and semi-log plots to discriminate between power-law and exponential behavior, respectively, of $C_e$ in the measured low-$T$ range, as shown in Fig.~\ref{Fig4}. 

Figures \ref{Fig4}(a) and \ref{Fig4}(c) clearly show that the low temperature data ($T < 0.3$ $T_c$) follow a power-law behavior with an exponent $m \approx 3.5-4$ for the $x = 0$ and $x = 0.04$ samples, respectively. This power-law behavior is typical of all samples with $x \leq 0.04$. While a power-law of $m = 3$ is consistent with point nodes, a power-law exponent $m > 3$ could be due to the fact that the gap is not purely nodal. Therefore, the fact that the low-$T$ electronic heat capacity follows a power-law $T$ dependence with a power of $m = 3.5-4$ indicates that the nodal gap is dominant for the samples with $x \leq 0.04$.  Our result is thus consistent with previously reported point nodes in PrPt$_4$Ge$_{12}$ \cite{Maisuradze2009}. 
We note that semi-log plots of the same low-$T$ data [Figs.~\ref{Fig4}(b) and \ref{Fig4}(d)] show that, indeed, these data do not follow an exponential behavior (the slopes of these plots decrease monotonically with decreasing $T$). 

On the other hand, Figs.~\ref{Fig4}(e) and ~\ref{Fig4}(f) show that the specific heat data from the $x=0.07$ sample displays exponential behavior. This exponential behavior is typical of all the samples with $x > 0.04$, and it seems to indicate that the nodeless gap is dominant for the samples with $x > 0.04$.

\begin{figure}
\centering
\includegraphics[width=\linewidth]{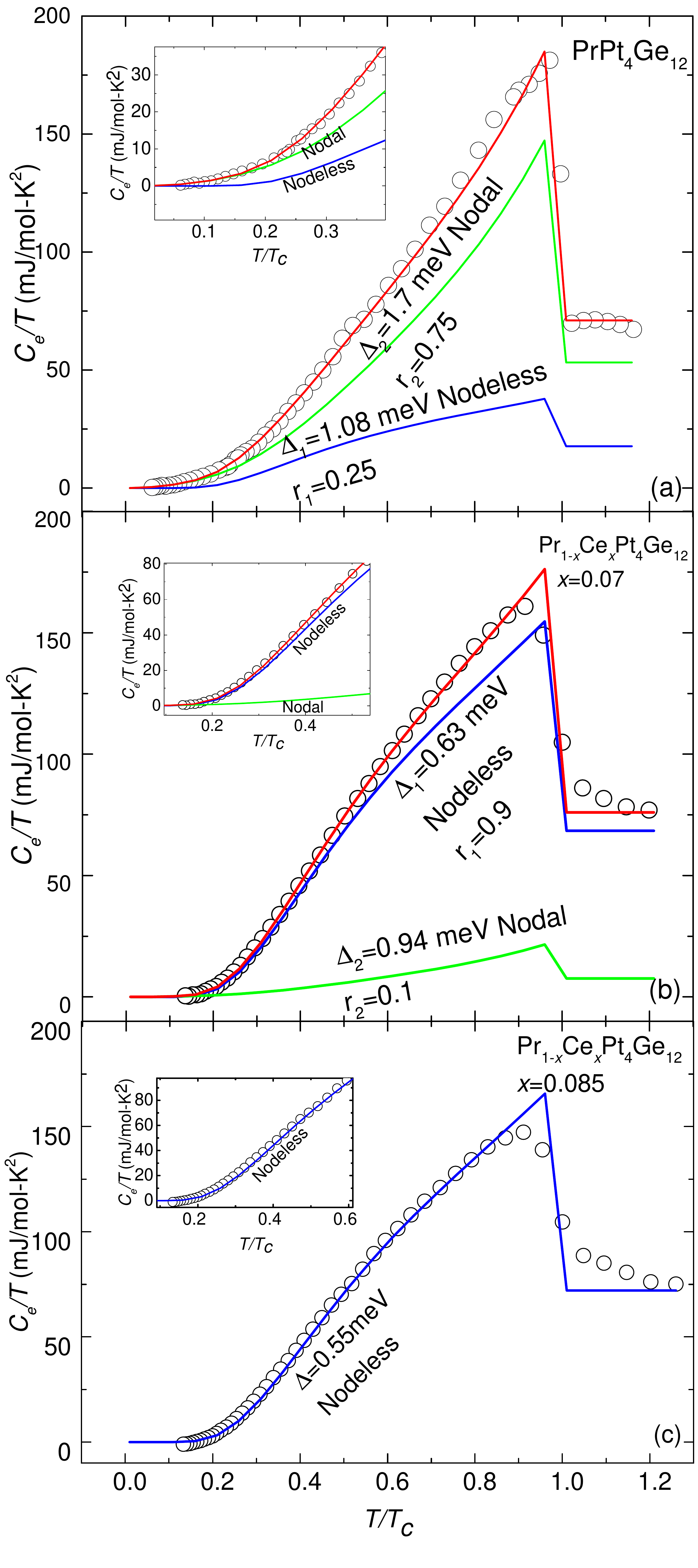}
\caption{ (Color online) Plot of the electronic specific heat $C_e/T$ vs $T/T_c$ for Pr$_{1-x}$Ce$_x$Pt$_4$Ge$_{12}$ with (a) $x = 0$ and (b) $x = 0.07$ using nodal and nodeless gaps and (c) $x=0.085$ with a nodeless gap. The red solid line gives the overall fit of the data, while the green and blue  lines are the individual contributions of the nodal and nodeless gaps, respectively. Insets: Low T region of the data in main panel. The details of the gap values are discussed in the text. }
\label{Fig5}
\end{figure}

In order to further explore the nature of the superconducting pairing, we used the following expression to evaluate the electronic specific heat $C_e$ in the superconducting state  \cite{Huang2006}:
\begin{equation}
\begin{split}
{C_e}&= 2\nu_F\beta k_B\frac{1}{4\pi}\int_0^{2\pi}{d}\phi\int_0^\pi\sin\theta {d}\theta\\ &\times\int_{-\infty}^{\infty}\left(-\frac{\partial f}{\partial E}\right)\left[E^2+\frac{1}{2}\frac{{d}\Delta^2}{{d}\beta}\right]{d}\epsilon, 
\end{split}  
\end{equation}
where $\nu_F$ is the density of states evaluated at the Fermi energy, $\beta = 1/k_BT$, $E=\sqrt{\epsilon^2+\Delta^2}$, $f=(1+e^{\beta E})^{-1}$, $\Delta=\Delta_0$ for isotropic $s$-wave pairing, and $\Delta=\Delta_0\sin n\theta$ for the case where the pairing wave function has point nodes. Based on this approach, we used a superposition of two different superconducting gaps $\Delta_{i}$ (\textit{i} = 1, 2), one nodal and one nodeless. Hence, there are two contributions to the electronic specific heat $C_e$ 
\begin{equation}\label{FitCe}
C_e(T) = r_1C_e(\Delta_{1}, T) + r_2C_e(\Delta_{2}, T), 
\end{equation} 
where $r_i\in[0,1]$ are weights for each contribution. 
We note that we subtracted the residual specific heat from the electronic contribution in order to be able to utilize the BCS model. We show the fit of the zero-field electronic specific heat data for Pr$_{1-x}$Ce$_x$Pt$_4$Ge$_{12}$ with a two band  model \cite{BouquetEPL2001}   in Figs.~\ref{Fig5}(a) and \ref{Fig5}(b) for the $x=0$ and $x=0.07$ samples, respectively. The data for the undoped sample are reproduced very well by using a smaller nodeless gap ($\Delta_{1} = 1.08$ meV) with a relative weight $r_1 = 0.25$ and a larger nodal gap with point nodes ($\Delta_{2} = 1.7$ meV) with a relative weight $r_2 = 0.75$ [Fig.~\ref{Fig5}(a)]. The data for the $x=0.07$ sample are reproduced very well by using an nodeless gap ($\Delta_{1} = 0.63$ meV) with a relative weight $r_1 = 0.9$, and a nodal gap ($\Delta_{2} = 0.94$ meV) with a relative weight $r_2 = 0.1$ [Fig.~\ref{Fig5}(b)]. 
The zero-field electronic specific heat data for Pr$_{1-x}$Ce$_x$Pt$_4$Ge$_{12}$ could be fitted very well with a larger nodal gap and a smaller nodeless gap for all samples with $0\leq x \leq 0.07$, implying that there are at least two Fermi surfaces in this doping range. Finally, for alloys with $x \geq 0.085$, the specific heat data could be fitted very well with only one nodeless gap as shown in Fig.~\ref{Fig5}(c) for the $x=0.085$ sample. 

\begin{figure}
\centering
\includegraphics[width=1\linewidth]{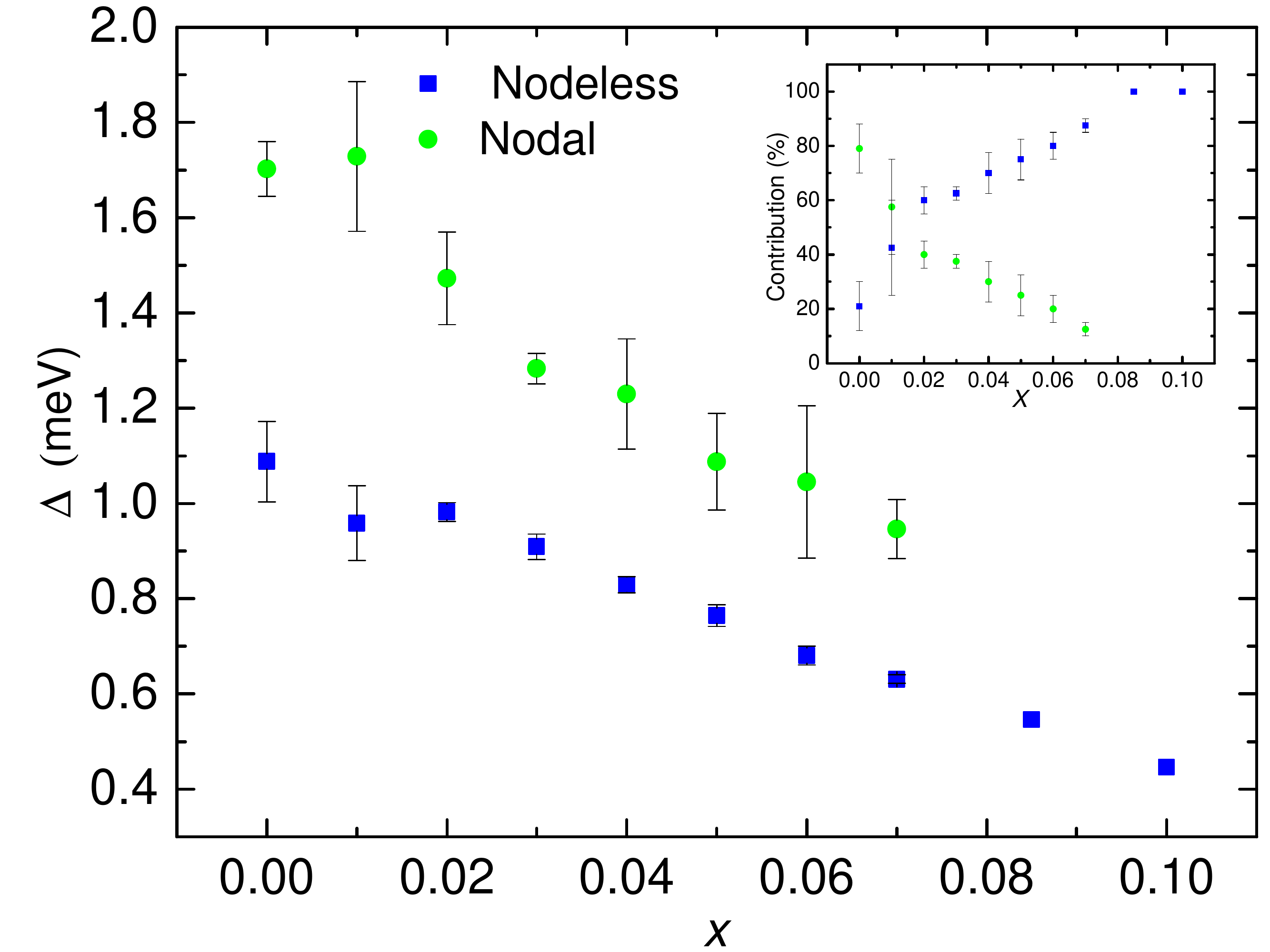}
\caption{(Color online) Gap values $\Delta$ plotted vs. Ce concentration $x$. The green solid circles are the nodal and the blue solid squares are the nodeless gap values used to reproduce the total heat capacity. Inset : Contribution of the nodal and nodeless gaps in the main panel plotted vs. Ce concentration.
\label{Fig6} }
\end{figure}

The doping dependence of the values of the two gaps is shown in Fig.~\ref{Fig6}. With increasing Ce concentration, the magnitude of the nodal gap decreases faster than that of the nodeless gap, consistent with the fact that the impurity scattering is more detrimental to the nodal gap than the nodeless gap. The inset of Fig.~\ref{Fig6} shows the doping dependence of the relative weight of the two gaps. Note that the contribution of the nodal gap decreases and that of the nodeless gap increases with increasing Ce concentration. Therefore, the density of states on the Fermi pocket with the nodal gap decreases with increasing Ce concentration and almost disappears for $x>0.07$. This is why the heat capacity data can be reproduced with only a single nodeless gap for samples with for $x > 0.07$. 

We reconcile the fact that the low-$T$ electronic heat capacity follows a power-law (exponential) behavior for all samples with $x \leq 0.04$ ($x > 0.04$) with the fact that all of the heat capacity data can be fitted very well with a two gap model - with one nodal and one nodeless gap - by drawing the reader's attention to the insets to Figs.~\ref{Fig5}. The inset to Fig.~\ref{Fig5}(a) shows that the nodeless gap is almost temperature independent in the low temperature region so that the overall contribution to the heat capacity is governed by the nodal gap in this temperature range; hence, consistent with the  power-law $T$ dependence of $C_e$ at low $T$ [Fig.~\ref{Fig4}(a)]. This behavior is typical for $0\leq x\leq 0.04$. 
On the other hand,  the inset to Fig.~\ref{Fig5}(b) shows that the nodal gap has a negligible temperature dependence for $0.04<x\leq 0.07$ so that the overall contribution to the heat capacity is governed by the nodeless gap in this temperature range; hence, the heat capacity follows exponential  $T$ dependence [Fig.~\ref{Fig4}(e)].   

\section{Conclusion}
In this paper, we analyzed the low-temperature specific heat data to investigate the nature of the superconducting order parameter in the Pr-based filled skutterudite system Pr$_{1-x}$Ce$_x$Pt$_4$Ge$_{12}$. Our findings indicate that specific heat has contributions from two Fermi pockets: one contribution originates from the pairing with a nodeless gap function, while the other one is best described by a model with a nodal gap function. The effect of Ce substitution is displayed in the monotonic suppression of the superconducting transition temperature. The larger gap remains nodal for Ce substitution below $x = 0.07$, while only the nodeless gap survives for $x > 0.07$. This property of Pr$_{1-x}$Ce$_x$Pt$_4$Ge$_{12}$ compounds resembles closely the multiband scenario of the La-substituted PrOs$_4$Sb$_{12}$ system, where the presence of both the nodal and the nodeless gap on different parts of Fermi surface has been suggested. Our results are also consistent with the recent findings of unconventional superconductivity suggested by the TRSB in the parent compound and its suppression through Ce doping. 

\section{Acknowledgments} 
The work at KSU was financially supported by the National Science Foundation grants DMR-1505826 (R.B.A., Y.P.S., S.Z. and C.C.A.) and DMR-1506547 (M.D.).  Materials synthesis and characterization at UCSD were supported by the US Department of Energy, Office of Basic Energy Sciences, Division of Materials Sciences and Engineering, under Grant No. DE-FG02-04ER46105.  Low temperature measurements at UCSD were sponsored by the National Science Foundation under Grant No. DMR 1206553. 

\bibliography{Ref_skutt}

\end{document}